# Beyond dpa: an atomistic framework for a quantitative description of radiation damage in $YBa_2Cu_3O_7$

Federico Ledda[1,2,3] | Daniele Torsello[*,1,2] | Davide Gambino[3,4] | Flyura Djurabekova[3] | Fabio Calzavara[1,2] | Niccolò Di Eugenio[1,2] | Ville Jantunen[3,5] | Antonio Trotta[6] | Erik Gallo[6] | Kai Nordlund[3] | Francesco Laviano[1,2]

[1]Department of Applied Science and Technology, Politecnico di Torino, Italy

[2]Istituto Nazionale di Fisica Nucleare, Sezione di Torino, Italy

[3]Department of Physics, University of Helsinki, Finland

[4]Department of Physics, Chemistry, and Biology, Linköping University, Sweden

[5]Culham Campus, United Kingdom Atomic Energy Authority, Abingdon,Oxfordshire, OX14 3DB, UK

[6]Eni S.p.A. ,Italy

**Correspondence**
* Corresponding author Daniele Torsello:
Email: daniele.torsello@polito.it

**Present address**
This is sample for present address text this is sample for present address text.

**Abstract**

Radiation damage in high-temperature cuprate superconductors represents one of the main technological challenges for their deployment in harsh environments, such as fusion reactors and accelerator facilities. Their complex crystal structure makes modeling irradiation effects in this class of materials a particularly demanding task, for which existing damage models remain inadequate. In this work, we develop an atomistic-based approach for describing primary radiation damage in $YBa_2Cu_3O_7$, by coupling Molecular Dynamics and Binary Collision Approximation simulations in a way that makes them complementary. When integrated with Primary Knock-on Atom spectra obtained from Monte Carlo codes, our results establish a framework for multiscale modeling of radiation damage, enabling quantitative estimates of several damage descriptors, such as defect production, defect clustering, and the effective damaged volume for any specific irradiation conditions where collision cascades dominate. This computational approach is suitable for the prediction of irradiation effects in any complex functional oxide, with applications ranging from aerospace to nuclear fusion and high-energy physics.

## 1 | INTRODUCTION

High-temperature superconductors (HTS), particularly the cuprate $YBa_2Cu_3O_{7-\delta}$ (YBCO), have emerged as leading candidates for efficient high-field magnets, as their technological maturity now enables operation far beyond the intrinsic field and temperature limits of conventional low-temperature superconductors[1,2,3,4]. These advances have established YBCO as a key enabling material for compact fusion reactors and next-generation high-energy accelerators, where superconductors must operate under intense radiation and extreme environmental conditions, bringing the radiation tolerance of such materials under the spotlight[5,6,7]. In harsh radiation environments, energetic particles inevitably interact with the crystal lattice, producing a broad variety of defects that can profoundly alter the functional properties of the material, like the critical current density $J_c$,[8] and the superconducting critical temperature, $T_c$[9]. Understanding how ionizing radiation affects YBCO is therefore crucial not only to predict its operational reliability, but also to unravel the microscopic mechanisms that link defect formation to changes in superconducting behavior[10]. To this end, irradiation studies with different particles, from electrons to light ions and fission neutrons, are being performed to emulate the complex damage environments expected in operation[11,12,13,14,15].

However, meaningful comparison between such experiments requires a common metric of radiation exposure: particle fluence alone is insufficient, as charge, mass, and energy determine distinct damage mechanisms. Traditionally, this role has been fulfilled by the Norget-Robinson-Torrens displacement-per-atom (NRT-dpa) concept, widely used to normalize irradiation effects in metallic alloys[18]. Originally developed for monatomic metals on the basis of early simulations, the model offers a fast estimate of damage levels and is routinely implemented in most Monte Carlo (MC) particle-transport codes. However, it should be regarded primarily as a measure of the energy deposited into the lattice, particularly when applied to complex compounds with mixed covalent and ionic bonding [19]. The limitations of the dpa formulation become evident when the degradation of superconducting properties is examined across irradiation studies employing different particle types. Comparisons among experiments with single crystals [16,15,17] employing fully penetrating particles (therefore able to eliminate extrinsic effects such as substrate stress [20,21]), reveal markedly different reductions in $T_c$ at comparable dpa values (an examples is shown in Fig. 1A). The discrepancy is particularly pronounced between electrons and fission neutrons, which differ greatly in charge, mass, and energy transfer per collision. Such divergence, which cannot be reconciled within the dpa formalism, points to the fundamentally different defect landscapes generated by each irradiation type and underscores the need for a more refined description of radiation damage, especially in functional materials.

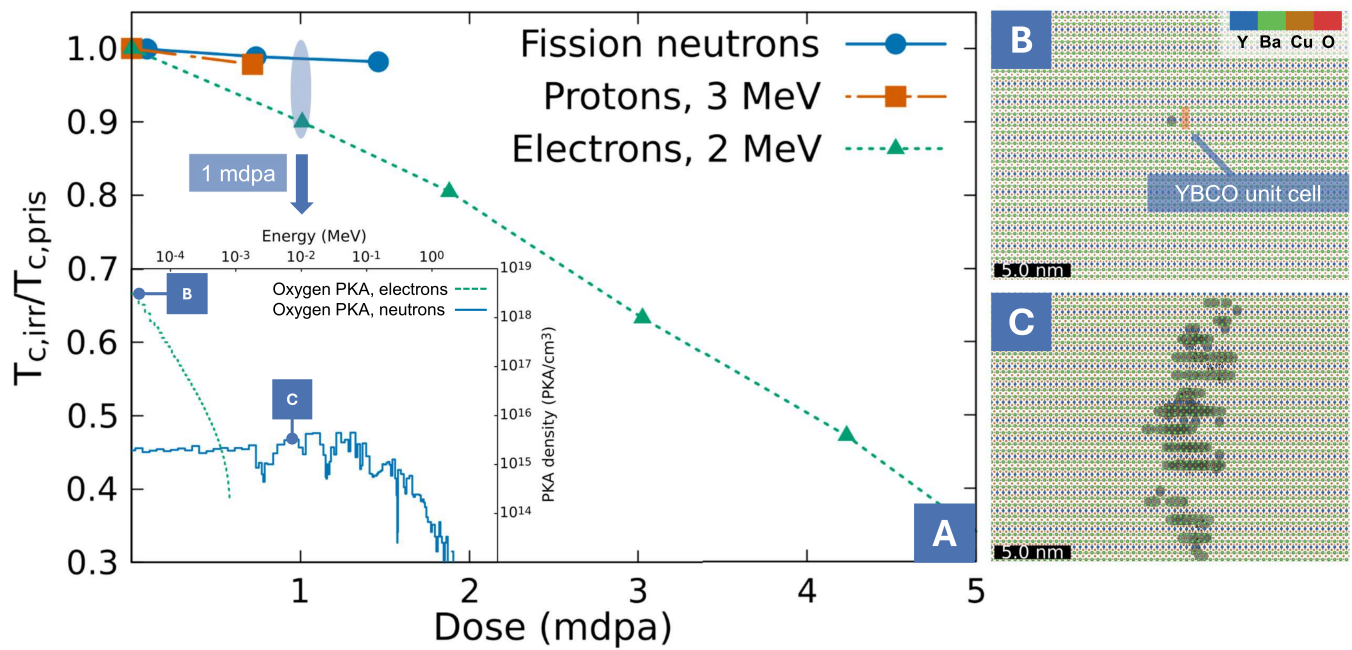

**FIGURE 1** A) Normalized $T_c$ of YBCO single crystals under three different irradiation conditions (protons [16], electrons [15], and fission neutrons [17]) plotted as a function of the dose (dpa). For the same dpa, markedly different reductions in $T_c$ are observed, indicating that dpa alone is insufficient to capture the extent of functionality degradation. The dpa values are estimated by the authors from the original fluence data and experiment descriptions. Oxygen primary knock-on atom spectra calculated for YBCO at a dose of 1 mdpa are presented in the inset, for electron (dotted line) and neutron (solid line) irradiation. Low-energy recoils (few eV) dominate under electron irradiation, whereas neutron irradiation produces high-energy recoils capable of generating extended defect structures. B) Example of defects produced by a 25 eV oxygen recoil at 20 K, obtained from MD simulations. Interstitial atoms are highlighted by a light-blue mesh. C) Example of a large defect cluster produced by a 7 keV oxygen recoil at 20 K, as obtained from MD simulations. Interstitial atoms are highlighted by a light-blue mesh.

In the irradiation regimes considered in this work, and relevant for most applications of functional materials [22,23,24,25], damage is governed by collision cascades initiated by the first atoms displaced by the incident particles, known as Primary Knock-on Atoms (PKAs) [26]. These atoms transfer their kinetic energy through successive collisions, generating cascades that ultimately define the defect population and distribution. Because the PKA energy sets the scale and morphology of these cascades, their spectra provide essential insight into the microscopic origin of irradiation effects and can be evaluated computationally, either through MC transport simulations for neutrons [27] or via the McKinley–Feshbach formalism for electrons [28]; computational details for the calculation are provided in the Supplementary Material. In YBCO, oxygen PKAs are particularly relevant, as they dominate statistically owing to the low displacement threshold [29]. When their spectra computed at the same dose level (1 mdpa) for the two example cases of Fig. 1 are compared (inset of Fig. 1A), the markedly different recoil-energy regimes of the two irradiation types become evident. Under electron irradiation, recoils of only a few eV prevail, whereas neutron irradiation produces PKAs with energies up to several MeV. Although particle-transport MC methods provide no direct structural information, the ensuing cascades can be explicitly modeled once the PKA energy is known [6] and a suitable interatomic potential [30] is available for Molecular Dynamics (MD) simulations. Recoils with energies only slightly above the displacement threshold, which largely dominate under electron irradiation, generate point-like defects in the form of a few Frenkel pairs, confined within a volume comparable to a single YBCO unit cell (Fig. 1B). In contrast, higher-energy recoils in the keV range, typical of neutrons irradiation, produce extended defect clusters and locally disordered regions spanning tens of unit cells (Fig. 1C). The relative abundance of these defect types can explain the different suppression of superconducting properties observed in cuprates, as they introduce scattering centers of different strength [31,32,12] (whose impact does not scale simply with size), yet such distinctions remain inaccessible to existing damage indicators. Despite their descriptive power, systematic MD simulations of collision cascades are computationally prohibitive and require significant expertise, preventing their routine use for experimental design or for identifying suitable damage proxies across irradiation conditions. A physically informed, multiscale and cuprate-specific atomistic description of radiation damage is therefore required to bridge the gap between the simplicity of dpa-based metrics and the accuracy of atomistic simulations. Motivated by these findings, we propose an approach that integrates an atomistic representation of defect formation into MC transport codes, providing a qualitative yet physically grounded description of the damage induced in YBCO.

## 2 | RESULTS AND DISCUSSION

### 2.1 | Method Overview

Constructing an atomistic description of radiation damage in YBCO requires, in practical terms, quantifying the response of the material to PKAs across the full energy spectrum relevant to irradiation (extending from few eV up to several MeV) and establishing a transferable dataset of analyzed cascades spanning the entire PKA-energy range. This dataset provides the atomistic foundation for coupling with MC particle-transport calculations to predict radiation damage under arbitrary conditions in which atomic displacements and collision cascades dominate, while retaining a computational cost and complexity comparable to the NRT model.

MD simulations offer a complete description of collision cascades, capturing many-body interactions and defect recombination, but their computational cost restricts the applicability of the method to relatively low recoil energies and small simulation volumes [33,34]. At the opposite extreme, Binary Collision Approximation (BCA) methods efficiently describe the high-energy ballistic phase of cascades, where atomic collisions are essentially binary and collective effects are negligible [34]. The separation between BCA and MD runs is also well motivated by that cascades induced by ions or recoils with energies exceeding the keV range are split into spatially separated subcascades [35]. Combining these two techniques therefore provides the most efficient route to achieve atomistic coverage over the entire energy spectrum [36].

On this basis, we introduce a hierarchical coupling strategy that combines the efficiency of BCA with the atomistic accuracy of MD. Cascades initiated by recoils with energies up to 2 keV (a threshold chosen based on physical and computational considerations, as detailed in the Supplementary Material) are resolved directly through MD in LAMMPS [37], capturing the full sequence of atomic rearrangements and defect recombination that takes place during the formation phase of the cascade. The resulting distributions of defects are compact, indicating that the damaged region can be accurately represented by a sphere centered along the PKA trajectory, with a radius equal to the gyration radius $R_g$(E) of the defect distribution (Fig. 2A, top). These simulations constitute a dataset for the low-energy regime, quantifying the number and type of Frenkel pairs and the characteristic spatial extent of the defected region for each PKA species. At higher energies, the early ballistic phase of the cascade is modeled using the BCA code CASWIN [38], transporting each recoil through successive collisions until all of them slow below the 2 keV threshold, thereby defining the backbone of the large cascades (Fig. 2A, bottom).

The high-energy backbone obtained from BCA is then coupled to the MD dataset to reconstruct the full cascade geometry. Each recoil within the MD energy window is regarded as the PKA of an individual sub-cascade and replaced by a virtual damage region centered along its trajectory, represented as a sphere whose radius corresponds to the $R_g$(E) derived from MD simulations at the same recoil energy. The defect yield associated with each recoil is likewise interpolated from the MD dataset. As long as the defect yield from individual recoils remains limited, so that the damaged volumes is not saturated, the interaction between sub-cascades can be neglected and their independent treatment provides an accurate approximation. The morphology of the reconstructed cascades is defined by the ensemble of these virtual regions and the associated defect yields (Fig. 2B).

With the cascade geometry reconstructed, the analysis focuses on the spatial organization and connectivity of the resulting damaged regions. (Fig. 2C). While the total number of produced defects, closely related to the overall disordered fraction, is a useful indicator of irradiation damage, it is insufficient on its own to predict superconducting property degradation in complex functional materials such as YBCO. The arrangement of defects is decisive for superconducting performance: extended clusters can, in some cases, enhance the $J_c$ by acting as effective pinning centers [39,40,41,8], whereas small isolated defects predominantly increase charge-carrier scattering and suppress $T_c$ and $J_c$ [12,42]. Assessing clustering is therefore essential, even though the definition of a "cluster" within a cascade is inherently dependent on the scale considered for the grouping criterion, dictated by the physical properties under investigation. Focusing on the superconducting behavior, we assume that the physically relevant length scale is the low temperature Cooper-pair coherence length, $\xi$(T=0) = 1.12 nm [43], which represents the spatial extent of the paired electronic state. Accordingly, clusters are defined as groups of defects separated by less than $2\xi$. In the reconstructed cascades this distance is measured between the surfaces of the spherical sub-cascade regions. The total volume affected by defects is obtained by computing the volume of each cluster using Gmsh [44], yielding a quantitative measure of the fraction of material structurally perturbed by irradiation. Thanks to this further analysis, clusters whose volume is smaller than that of a sphere with radius $\xi$ are in this work classified as *small clusters*, representing isolated defect configurations most relevant for charge-carrier scattering.

With this hierarchical coupling approach, excellent agreement is obtained with respect to full MD simulations of large cascades, confirming the validity of the scheme. The detailed procedure and the verification of these assumptions are provided in the Supplementary Materials.

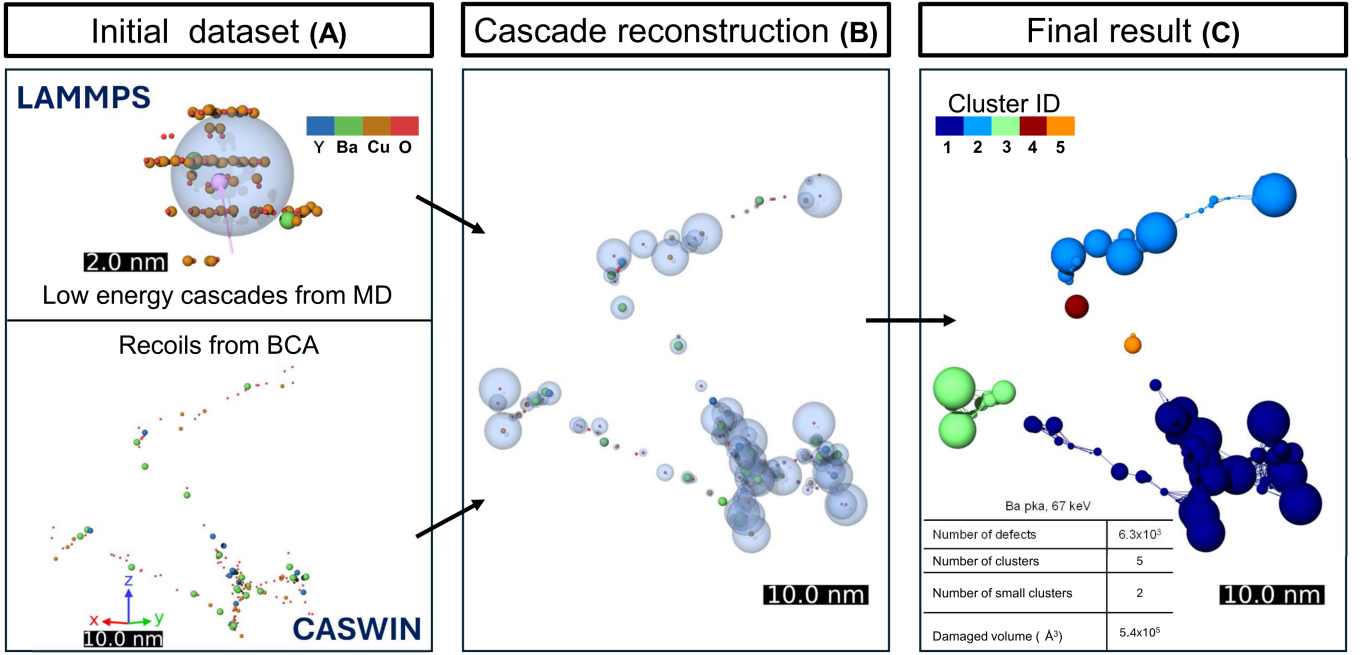

**FIGURE 2** (A) Initial datasets: Top —example of a low-energy cascade (Ba PKA, 245 eV) from MD simulations, representative of the database of cascades computed up to 2 keV for all atomic species. The defected region is well approximated by a sphere centered along the PKA trajectory with a radius equal to the $R_g$ of the defect distribution. Bottom — example of recoil distribution from a high-energy (60 keV) Ba cascade obtained with the BCA code CASWIN[38], illustrating the backbone of large cascades. (B) Cascade reconstruction: each BCA recoil within the MD energy window (E < 2 keV) is replaced by a virtual spherical damage region whose radius $R_g(E)$ and defect yield interpolated from the MD dataset. (C) Final reconstructed cascade and clustering analysis. Clusters are identified by connecting spheres whose surfaces are separated by less than 2ξ (ξ ≈ 1 nm), corresponding to the superconducting coherence length in YBCO. The reconstructed geometry can also be exported to CAD format to estimate the total damaged volume using Gmsh[44].

The resulting dataset quantifies novel damage descriptors such as the *defect yield* (Fig. 3A), *damaged volume* (Fig. 3B), and *clustering statistics* (Fig. 3C,D) for each PKA species across the full recoil-energy spectrum. The sublinear increase in defect production with energy reflects the progressive reduction in damage efficiency as cascades expand and energy dissipation becomes increasingly non-local. The scaling with PKA mass originates from momentum transfer, with heavier atoms inducing larger and denser cascades. In contrast, oxygen PKAs exhibit a characteristic saturation beyond ≈ 0.5 MeV, where limited momentum transfer collisions and enhanced electronic stopping restrict further defect generation.

Clustering analysis reveals that cascades initiated by Y, Ba, and Cu PKAs share similar morphologies, while those generated by oxygen are markedly more fragmented, producing a higher density of small, spatially isolated clusters. This behavior mirrors the defect-volume scaling, with heavy atoms forming compact regions of extensive disorder, whereas oxygen generates numerous confined defects that occupy a smaller overall volume.

## 2.2 | Application to representative irradiation cases

| Quantity | Neutron irradiation | Electron irradiation |
|---|---|---|
| Nominal dose (NRT-mdpa) | 1 | 1 |
| Frenkel pair density (cm$^{-3}$) | $5.48 \times 10^{20}$ | $2.45 \times 10^{20}$ |
| Cluster density (cm$^{-3}$) | $1.44 \times 10^{18}$ | $3.45 \times 10^{19}$ |
| Small-cluster fraction (%) | 53 | 99 |
| Mean defects per cluster | 405 | 7 |
| Damaged volume fraction (%) | 0.98 | 1.22 |

**TABLE 1** Predicted radiation-damage metrics for YBCO under equivalent nominal doses (1 mdpa). Values are obtained from the coupled MD–BCA dataset convolved with the respective PKA spectra. Despite identical doses, neutron irradiation produces fewer but larger and denser clusters, whereas electron irradiation yields a higher density of small, spatially isolated defects.

Having established the atomistic response of YBCO to individual recoils, the framework can now be applied to realistic irradiation scenarios to quantify the resulting damage landscape. In particular, electron and fission-neutron irradiations, each corresponding to an accumulated dose of 1 mdpa as discussed in the introduction, provide an ideal benchmark to test the model and to elucidate the distinct degradation behaviors that the NRT-dpa framework fails to capture.

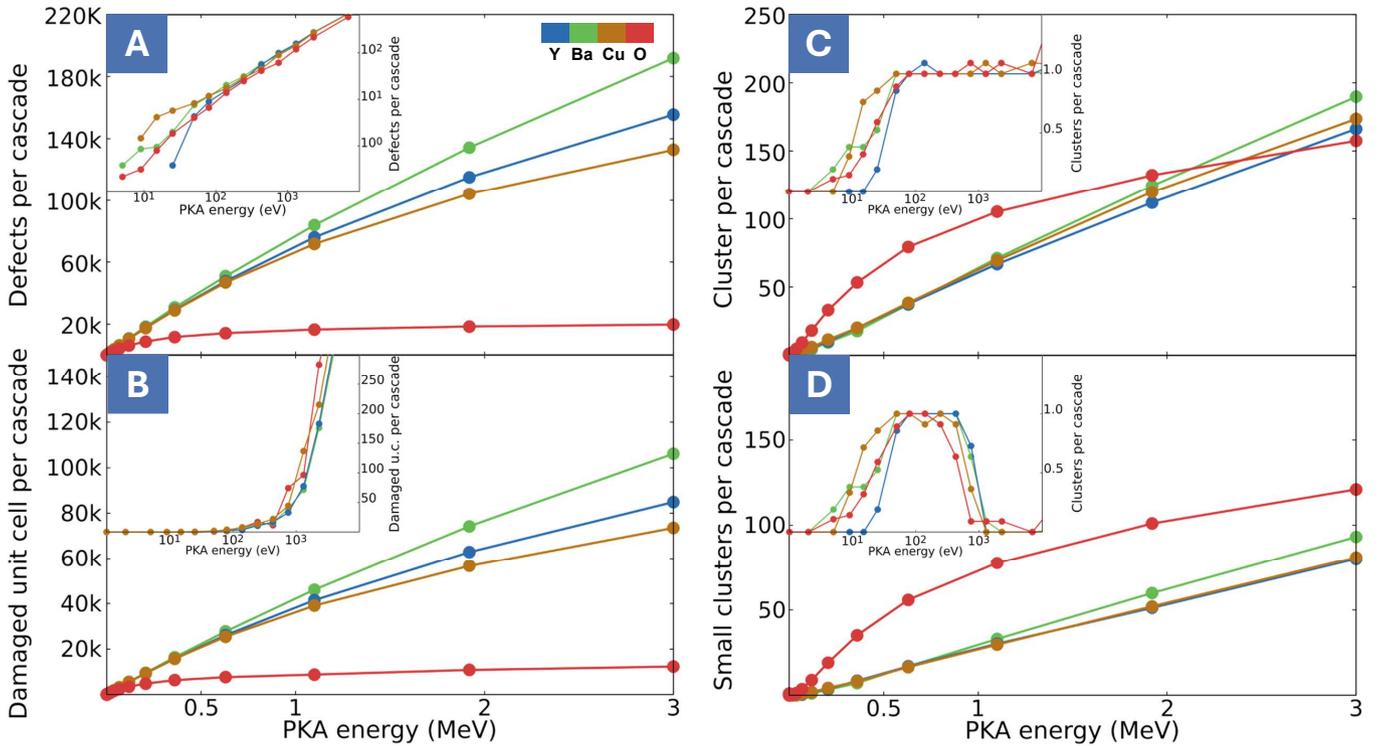

**FIGURE 3** Defect yield (A) and total damaged volume (B) per cascade as a function of PKA energy for each atomic species at 20 K. The sublinear scaling reflects the decreasing efficiency of defect production with increasing cascade size, while heavier PKAs transfer momentum more effectively, generating denser damage. Number of clusters (C) and small clusters (D), defined as defect regions with a volume smaller than that of a sphere of radius ξ ≈ 1 nm) per cascade. Y, Ba, and Cu cascades exhibit comparable clustering behaviour, whereas O-induced cascades produce a markedly higher fraction of small, spatially isolated clusters. Insets in all panels highlight the low-energy regime resolved by direct MD simulations. The non-monotonic trend observed for the small-cluster counts originates from a sorting effect: at low energies, nearly all clusters are small, leading to an initial rise; as cascades merge into larger defect regions, the number of small clusters decreases, before increasing again when fragmentation into subcascades appears at higher energies.

Although the two irradiation scenarios correspond to identical nominal doses, the predicted defect landscapes differ markedly (as detailed by our results shown in Tab 1): the total number of Frenkel pairs is comparable, as expected for equal NRT-dpa values, yet the model resolves a pronounced divergence in the resulting defect morphology. Electron irradiation, dominated by low-energy oxygen recoils (inset Fig 1A), produces a high density of small, spatially isolated clusters, each typically containing fewer than ten defects. Neutron irradiation, by contrast, generates energetic PKAs, leading to extended cascades and dense regions comprising hundreds of Frenkel pairs. These model predictions are fully consistent with well-established experimental observations and allow a quantitative treatment of the phenomenon, allowing us to capture the distinct defect landscapes arising from different irradiation types, providing information far richer than a single dose metric and offering a physically grounded basis for designing and interpreting irradiation experiments.

## 3 | CONCLUSION

In conclusion, this work demonstrates the intrinsic limitations of simple dose-based metrics such as the NRT-dpa when applied to complex functional materials, and introduces a multiscale computational framework that couples MD with the BCA to provide a physically grounded description of radiation damage, using YBCO as a case study. The approach overcomes the intrinsic restrictions of pure MD, limited to the keV energy range and small supercells, by leveraging BCA to access the high-energy regime. The resulting MD–BCA dataset establishes a transferable link between atomistic physics and engineering-scale predictions, offering detailed structural insight at a computational cost comparable to that of routine MC calculations and far beyond the descriptive power of standard NRT-dpa models. When applied to realistic irradiation conditions, the framework captures the distinct defect morphologies induced by different particle types, enabling quantitative characterization of the damaged landscape through complementary metrics such as *defect yield*, *cluster density*, and *damaged-volume fraction*. This capability provides a powerful interpretative tool for existing experiments and a predictive basis for the design of future irradiation studies. While the present framework offers a detailed atomistic picture of radiation-induced disorder, establishing direct links between these structural changes and the evolution of superconducting properties remains an open challenge. Progress in this direction will rely on expanding the experimental basis, particularly through systematic studies on single crystals under well-controlled irradiation conditions, together with continued theoretical efforts aimed at connecting the microscopic defect landscape to macroscopic functional behavior. In parallel, extending the computational analysis to longer timescales will be essential to capture defect migration, recovery processes, and the long-term evolution of the irradiated microstructure.

## SUPPORTING INFORMATION

Supporting Information is available from the Wiley Online Library or from the author.

## ACKNOWLEDGMENTS

FedL acknowledges that this publication is part of the project PNRR-NGEU which has received funding from the MUR – DM 352/2022. NDE acknowledges that this publication is part of the project PNRR-NGEU which has received funding from the MUR – DM 117/2023 (or DM 118/2023). Support from Eni S.p.A. is acknowledged. F.D. and V.J. acknowledge the Research Council of Finland project SPATEC (Grant No 349690) for financial support. This work has been partly carried out within the framework of the EUROfusion Consortium, funded by the European Union via the Euratom Research and Training Programme (Grant Agreement No 101052200 — EUROfusion). Views and opinions expressed are how ever those of the author(s) only and do not necessarily reflect those of the European Union or the European Commission. Neither the European Union nor the European Commission can be held responsible for them. High-Performance Computing resources were provided by CINECA (Italy), JFRS-1 (Japan), and the Swedish National Infrastructure for Computing (SNIC) at the PDC Center for High Performance Computing (Dardel) and the National Supercomputer Centre (Tetralith). The authors also acknowledge the technical support teams of these facilities for their assistance.

## CONFLICT OF INTEREST

The authors declare no potential conflict of interests.

## DATA AVAILABILITY STATEMENT

The data that support the findings of this study are available from the corresponding author upon reasonable request.

## REFERENCES

1. Uglietti D. A review of commercial high temperature superconducting materials for large magnets: from wires and tapes to cables and conductors. *Superconductor Science and Technology*. 2019;32(5):053001. doi: 10.1088/1361-6668/ab06a2
2. Molodyk A, Samoilenkov S, Markelov A, et al. Development and large volume production of extremely high current density YBa2Cu3O7 superconducting wires for fusion. *Scientific reports*. 2021;11(2084):1–11. doi: 10.1038/s41598-021-81559-z
3. Godeke A. High temperature superconductors for commercial magnets. *Superconductor Science and Technology*. 2023;36(11):113001. doi: 10.1088/1361-6668/acf901
4. Huang R, Chen J, Liu Z, Wang G, Cai C. Significantly Improving the Flux Pinning of YBa2Cu3O7-δ Superconducting Coated Conductors via BaHfO3 Nanocrystal Addition Using Multistep Film Growth Method. *Advanced Functional Materials*. 2024;34(36):2401251. doi: https://doi.org/10.1002/adfm.202401251
5. Sorbom B, Ball J, Palmer T, et al. ARC: A compact, high-field, fusion nuclear science facility and demonstration power plant with demountable magnets. *Fusion Engineering and Design*. 2015;100:378–405.
6. Torsello D, Gambino D, Gozzelino L, Trotta A, Laviano F. Expected radiation environment and damage for YBCO tapes in compact fusion reactors. *Superconductor Science and Technology*. 2022;36(1):014003. doi: 10.1088/1361-6668/aca369
7. Iliffe W, Chislett-McDonald S, Harden F, et al. Progress in the STEP Programme Toward Understanding REBCO Coated Conductors in the Fusion Environment. *IEEE Transactions on Applied Superconductivity*. 2025;35(5):1-5. doi: 10.1109/TASC.2024.3523248
8. Ruiz H, Hänisch J, Polichetti M, et al. Critical current density in advanced superconductors. *Progress in Materials Science*. 2026;155:101492. doi: https://doi.org/10.1016/j.pmatsci.2025.101492
9. Rullier-Albenque F, Vieillefond PA, Alloul H, Tyler AW, Lejay P, Marucco JF. Universal Tc depression by irradiation defects in underdoped and overdoped cuprates?. *Europhysics Letters*. 2000;50(1):81. doi: 10.1209/epl/i2000-00238-x
10. Alloul H, Bobroff J, Gabay M, Hirschfeld PJ. Defects in correlated metals and superconductors. *Rev. Mod. Phys.*. 2009;81:45–108. doi: 10.1103/RevModPhys.81.45
11. Fischer DX, Prokopec R, Emhofer J, Eisterer M. The effect of fast neutron irradiation on the superconducting properties of REBCO coated conductors with and without artificial pinning centers. *Superconductor Science and Technology*. 2018;31(4):044006. doi: 10.1088/1361-6668/aaadf2
12. Unterrainer R, Gambino D, Semper F, et al. Responsibility of small defects for the low radiation tolerance of coated conductors. *Superconductor Science and Technology*. 2024;37(10):105008. doi: 10.1088/1361-6668/ad70db
13. Nicholls RJ, Diaz-Moreno S, Iliffe W, et al. Understanding irradiation damage in high-temperature superconductors for fusion reactors using high resolution X-ray absorption spectroscopy. *Communications Materials*. 2022;3(1):1–14. Publisher: Nature Publishing Groupdoi: 10.1038/s43246-022-00272-0
14. Iiffe W, Adams K, Peng N, et al. The effect of in situ irradiation on the superconducting performance of REBa$_2$Cu$_3$O$_{7-\delta}$-coated

conductors. *MRS Bulletin*. 2023;48:710. doi: 10.1557/s43577-022-00473-5

15. Solovjov AL, Rogacki K, Shytov NV, et al. Influence of strong electron irradiation on fluctuation conductivity and pseudogap in $YBa_2Cu_3O_{7−δ}$ single crystals. *Phys. Rev. B.* 2025;111:174508. doi: 10.1103/PhysRevB.111.174508

16. Civale L, Marwick AD, McElfresh MW, et al. Defect independence of the irreversibility line in proton-irradiated Y-Ba-Cu-O crystals. *Phys. Rev. Lett.*. 1990;65:1164–1167. doi: 10.1103/PhysRevLett.65.1164

17. Sauerzopf FM, Wiesinger HP, Weber HW, Crabtree GW. Analysis of pinning effects in $YBa_2Cu_3O_{7−δ}$ single crystals after fast neutron irradiation. *Phys. Rev. B.* 1995;51:6002–6012. doi: 10.1103/PhysRevB.51.6002

18. Norgett M, Robinson M, Torrens I. A proposed method of calculating displacement dose rates. *Nuclear Engineering and Design*. 1975;33(1):50-54. doi: https://doi.org/10.1016/0029-5493(75)90035-7

19. Nordlund K, Zinkle SJ, Sand AE, et al. Improving atomic displacement and replacement calculations with physically realistic damage models. *Nature Communications*. 2018;9. doi: 10.1038/s41467-018-03415-5

20. Torsello D, Celentano G, Civale L, et al. Roadmap for the investigation of irradiation effects in HTS for fusion. *Superconductor Science and Technology*. 2025;38(5):053501. doi: 10.1088/1361-6668/adce40

21. Telkhozhayeva M, Girshevitz O. Roadmap toward Controlled Ion Beam-Induced Defects in 2D Materials. *Advanced Functional Materials*. 2024;34(45):2404615. doi: https://doi.org/10.1002/adfm.202404615

22. Thiruraman JP, Masih Das P, Drndić M. Irradiation of Transition Metal Dichalcogenides Using a Focused Ion Beam: Controlled Single-Atom Defect Creation. *Advanced Functional Materials*. 2019;29(52):1904668. doi: https://doi.org/10.1002/adfm.201904668

23. Parkhomenko HP, Solovan MM, Sahare S, et al. Impact of a Short-Pulse High-Intense Proton Irradiation on High-Performance Perovskite Solar Cells. *Advanced Functional Materials*. 2024;34(10):2310404. doi: https://doi.org/10.1002/adfm.202310404

24. Li G, Zou C, Wang F, et al. Atomic-Precision Manipulation of Defects in RuO2 Nanocrystals via Electron-Beam. *Advanced Functional Materials*. 2024;34(51):2410524. doi: https://doi.org/10.1002/adfm.202410524

25. An B, Deng Y, Jin Z, Sun S. Scintillators for Neutron Detection and Imaging: Advances and Prospects. *Advanced Functional Materials*. 2025;35(19):2422522. doi: https://doi.org/10.1002/adfm.202422522

26. Nordlund K, Zinkle SJ, Sand AE, et al. Primary radiation damage: A review of current understanding and models. *Journal of Nuclear Materials*. 2018;512:450-479. doi: https://doi.org/10.1016/j.jnucmat.2018.10.027

27. Sato T, Iwamoto Y, Hashimoto S, et al. Recent improvements of the particle and heavy ion transport code system – PHITS version 3.33. *Journal of Nuclear Science and Technology*. 2024;61(1):127–135. doi: 10.1080/00223131.2023.2275736

28. Lucasson PG, Walker RM. Production and Recovery of Electron-Induced Radiation Damage in a Number of Metals. *Phys. Rev.*. 1962;127:485–500. doi: 10.1103/PhysRev.127.485

29. Ledda F, Torsello D, Pettinari D, et al. 3D Neutronic and Secondary Particles Analysis on $YBa_2Cu_3O_{7−δ}$ Tapes for Compact Fusion Reactors. *IEEE Transactions on Applied Superconductivity*. 2024;34(3):1-5. doi: 10.1109/TASC.2024.3379114

30. Gray RL, Rushton MJD, Murphy ST. Molecular dynamics simulations of radiation damage in YBa2Cu3O7. *Superconductor Science and Technology*. 2022;35(3):035010. doi: 10.1088/1361-6668/ac47dc

31. Wang LL, Hirschfeld PJ, Cheng HP. Ab initio calculation of impurity effects in copper oxide materials. *Phys. Rev. B.* 2005;72:224516. doi: 10.1103/PhysRevB.72.224516

32. Özdemir HU, Mishra V, Lee-Hone NR, et al. Effect of realistic out-of-plane dopant potentials on the superfluid density of overdoped cuprates. *Phys. Rev. B.* 2022;106:184510. doi: 10.1103/PhysRevB.106.184510

33. Roy A, Nandipati G, Casella AM, Senor DJ, Devanathan R, Soulami A. A review of displacement cascade simulations using molecular dynamics emphasizing interatomic potentials for TP-BAR components. *npj Materials Degradation*. 2025;9(1):1. doi: 10.1038/s41529-024-00536-9

34. Heinisch HL. Defect production in high energy cascades: the roles of molecular dynamics and binary collision simulations. *Radiation Effects and Defects in Solids*. 1994;129(1-2):113–116. doi: 10.1080/10420159408228887

35. De Backer A, Sand AE, Nordlund K, Luneville L, Simeone D, Dudarev SL. Subcascade formation and defect cluster size scaling in high-energy collision events in metals. *Europhysics Letters*. 2016;115(2):26001. doi: 10.1209/0295-5075/115/26001

36. Ortiz CJ. A combined BCA-MD method with adaptive volume to simulate high-energy atomic-collision cascades in solids under irradiation. *Computational Materials Science*. 2018;154:325-334. doi: https://doi.org/10.1016/j.commatsci.2018.07.058

37. Thompson AP, Aktulga HM, Berger R, et al. LAMMPS - a flexible simulation tool for particle-based materials modeling at the atomic, meso, and continuum scales. *Comp. Phys. Comm.*. 2022;271:108171. doi: 10.1016/j.cpc.2021.108171

38. Pugacheva TS, others . *Nuclear Instruments and Methods in Physics Research Section B: Beam Interactions with Materials and Atoms*. 1998;141:99–104.

39. Puig T, Obradors X, Martínez B, Sandiumenge F, O'Callaghan J, Rabier J. Direct identification of extended defects as vortex pinning centers in melt-textured $YBa_2Cu_3O_7$-$Y_2BaCuO_5$ composites. *IEEE Transactions on Applied Superconductivity*. 1999;9(2):2663–2666. doi: 10.1109/77.785034

40. Mukaida M, Matsumoto K, Horide T, Ichinose A, Yoshida Y, Horii S. Structural and magnetotransport properties of YBa2Cu3O7−δ films exhibiting enhanced flux pinning. *Japanese Journal of Applied Physics, Part 2*. 2005;44:L246–L? doi: 10.1143/JJAP.44.L246

41. Obradors X, Puig T. Pin the vortex on the superconductor. *nature materials*. 2024;23(10):1311–1312. doi: 10.1038/s41563-024-01990-1

42. Eisterer M, Bodenseher A, Unterrainer R. Degradation of superconductors due to impurity scattering: predicting the performance loss in fusion magnets. *Superconductor Science and Technology*. 2025;38(10):10LT01. doi: 10.1088/1361-6668/ae1553

43. Ando Y, Segawa K. Magnetoresistance of Untwinned $YBa_2Cu_3O_y$ Single Crystals in a Wide Range of Doping: Anomalous Hole-Doping Dependence of the Coherence Length. *Phys. Rev. Lett.*. 2002;88:167005. doi: 10.1103/PhysRevLett.88.167005

44. Geuzaine C, Remacle JF. Gmsh: A three-dimensional finite element mesh generator with built-in pre- and post-processing facilities. *International Journal for Numerical Methods in Engineering*. 2009;79(11):1309–1331. doi: 10.1002/nme.2579